\documentclass[reprint, amsmath,amssymb, aps, pra]{revtex4-1}

\usepackage{graphicx}
\usepackage{dcolumn}
\usepackage{bm}
\usepackage{subcaption} 

\begin{document}

\title{ Scalar dark matter in the radio-frequency band: atomic-spectroscopy search results}

\author{D. Antypas}

\affiliation{Helmholtz-Institut Mainz, Johannes Gutenberg-Universit{\"a}t Mainz, Mainz 55128, Germany}

\author{O. Tretiak}
\affiliation{Helmholtz-Institut Mainz, Johannes Gutenberg-Universit{\"a}t Mainz, Mainz 55128, Germany}

\author{A. Garcon}
\affiliation{Helmholtz-Institut Mainz, Johannes Gutenberg-Universit{\"a}t Mainz, Mainz 55128, Germany}

\author{R. Ozeri}
\affiliation{Department of Physics of Complex Systems, Weizmann Institute of Science, Rehovot, Israel 7610001}

\author{G. Perez }
\affiliation{Department of Particle Physics and Astrophysics, Weizmann Institute of Science, Rehovot, Israel 7610001}

\author{D. Budker}
\affiliation{Helmholtz-Institut Mainz, Johannes Gutenberg-Universit{\"a}t Mainz, Mainz 55128, Germany}
\affiliation{Department of Physics, University of California at Berkeley, California 94720-300, USA}

\date{\today}

\begin{abstract}

Among the prominent candidates for dark matter are bosonic fields with small scalar couplings to the Standard-Model particles. Several techniques are employed to search for such couplings and the current best constraints are derived from tests of gravity or atomic probes. In experiments employing atoms, observables would arise from expected dark-matter-induced oscillations in the fundamental constants of nature. These studies are primarily sensitive to underlying particle masses below \mbox{$10^{-14}$ eV}. We present a method to search for fast oscillations of fundamental constants using atomic spectroscopy in cesium vapor. We demonstrate sensitivity to scalar interactions of dark matter associated with a  particle mass in the range  $8\cdot10^{-11}$ to $4\cdot 10^{-7}$ eV. In this range our experiment yields constraints on such interactions, which within the framework of an astronomical-size dark matter structure, are comparable with, or better than, those provided by experiments probing deviations from the law of gravity.


\end{abstract}

\pacs{Valid PACS appear here}

\maketitle

\textit{Introduction} - The fundamental constants (FC) of nature are invariant in time within the Standard Model (SM) of particle physics, but become dynamical in a number of theories beyond the SM. This possibility has motivated diverse studies that constrain both present-day FC drifts, and changes of FC between the present time and an earlier time in the universe (see, for example, reviews \cite{Safronova2018SearchMolecules, Uzan2015TheConstants} and references therein). 

The FC are expected to oscillate in cases where the SM fields couple to an ultra-light scalar field, coherently oscillating to account for the observed dark matter (DM) density. 
Within this class of models, FC oscillations are expected to occur at the Compton frequency of the DM particle, $\omega_C=m_{\phi}$ \footnote{We use natural units, where $\hbar=c=1$.}, where $m_{\phi}$ is the particle's mass. 
Such a scenario is particularly motivated in two main cases: (i) where the DM candidate is a dilaton, and its coupling to the SM particles is dictated by scale invariance \cite{Arvanitaki2015SearchingClocks, Graham2015ExperimentalParticles}; and (ii) the DM is the relaxion field \cite{Graham2015CosmologicalScale}, dynamically misaligned from its local minima \cite{Banerjee2018CoherentMatter}, and its coupling to the SM fields arises due to its mixing with the Higgs \cite{Flacke2017PhenomenologyMixing}.  Relaxions may form gravitationally bound objects \cite{Banerjee2019RelaxionPhysics}, thereby increasing the local DM density and enhancing the observability of the scenario.  

There are several proposed schemes to probe light DM that has scalar couplings to SM matter. These include suggestions to look for variations of the fine-structure constant $\alpha$ using atomic clocks \cite{ Stadnik2015SearchingInterferometry, Stadnik2016EnhancedDetection, Arvanitaki2015SearchingClocks,Safronova2018TwoConstant, Savalle2019NovelClocks}, as well as variations in the length of solid objects \cite{Stadnik2015SearchingInterferometry, Stadnik2016EnhancedDetection, Arvanitaki2016SoundDetectors,Geraci2018SearchingCavities} which would arise from oscillations in $\alpha$ or the electron mass $m_e$. Direct detection of light scalar DM by probing a DM-induced equivalence-principle (EP)-violating force has been also suggested \cite{Graham2016DarkAccelerometers, Hees2018ViolationMatter}. Existing limits on scalar DM-SM matter interactions come from astrophysical considerations  \cite{Stadnik2015CanNature} as well as table-top experiments including radio-frequency (rf) spectroscopy in atomic Dy  \cite{VanTilburg2015SearchSpectroscopy,Stadnik2016ImprovedSpectroscopy,Leefer2016SearchParticles}, long-term comparison of Cs and Rb clocks \cite{ Hees2016SearchingComparisons,Stadnik2016ImprovedSpectroscopy}, a network of atomic clocks \cite{Wcisoetal.2018NewClocks},  EP and fifth-force (FF) experiments  \cite{Toubouletal.2017MICROSCOPEPrinciple, Berge2018MICROSCOPEDilaton, Adelberger2009TorsionPhysics, Wagner2012Torsion-balancePrinciple}. In an ongoing experiment \cite{Kennedy2018J.YeSr}, a comparison of a Sr atomic clock with a Si cavity \cite{Stadnik2016EnhancedDetection} is employed to probe scalar DM couplings at frequencies up to 10 Hz, while in \cite{Aharony2019ConstrainingDecoupling}, a similar scheme involving spectroscopy with a single Sr$^{+}$ ion was used to probe the 1 Hz-1 MHz region (mass range $4\cdot10^{-15}-4\cdot10^{-9}$ eV).  

 The most stringent bounds to-date on scalar DM-SM matter couplings are the result of searches with atomic probes in the regime below the Hz-level, or tests of gravity in EP and FF apparatus. The latter experiments provide constraints up to a frequency of $\sim$ $10^9$ Hz.  Here we present an atomic spectroscopy method for detection of rapid variations of $\alpha$ and $m_e$ that extends the frequency range probed well into the rf-band, up to 100 MHz. The rf-band is already accessible by EP/FF searches indirectly. Direct probing of fast FC variations with atoms, however, offers a conceptually different approach to studying scalar DM in the particular regime. As we will see, the rf-frequency regime is especially interesting for searches for FC oscillations associated with the presence of astronomical-size DM objects. 
 
 Our method involves a search for oscillations in the energy spacing $\Delta E$ between two electronic levels in atomic $^{133}$Cs, the ground state 6s$^2$S$_{1/2}$ and excited state 6p$^2$P$_{3/2}$, while the corresponding  transition is resonantly excited with continuous-wave laser light of frequency $f_L\approx f_{atom}$, where $f_{atom}\equiv\Delta E/2\pi$. As this level spacing is approximately  proportional \footnote{The dependence $\Delta E\propto m_e\alpha^2$ is accurate to within relativistic corrections of order $(\alpha Z)^2 \approx$ 16\% in Cs} to the Rydberg constant $ \propto m_e\alpha^2$ \cite{Dzuba1999Space-timeAtoms, Dzuba1999CalculationsConstants}, our scheme is sensitive to oscillations of $m_e$ and $\alpha$. 

In the presence of a light scalar DM field $\phi$, $\alpha$ and $m_e$ acquire an oscillatory component, induced by the oscillations of the field at the Compton frequency of the DM particle. On time scales shorter than its coherence time, this field can be expressed as \cite{Banerjee2018CoherentMatter}:
\begin{equation}
\label{eq:phi}
\phi(\vec{r},t)\approx\frac{\sqrt{2\rho_{\rm DM}}}{m_{\phi}}\sin({m_{\phi}t}),
\end{equation}
where $\rho_{\rm DM}\approx0.4 $ GeV$/cm^3$ is the local DM density \cite{Catena2010ADensity}. The quantities $\alpha$ and $m_e$ follow the oscillations of $\phi$ and can be written as:
\begin{equation}
\label{eq:alphaVar}
\alpha(\vec{r},t)=\alpha_0\big[1+g_{\gamma}\phi(\vec{r},t)\big],
\end{equation}
\begin{equation}
\label{eq:mVar}
m_e(\vec{r},t)=m_{e,0}\Big[1+\frac{g_e}{m_{e,0}}\phi(\vec{r},t)\Big],
\end{equation}
where and $g_{\gamma}$, $g_{e}$ are coupling constants of DM to the photon and the electron, respectively. These are assumed independent for a generic DM candidate, but are related within the relaxion DM scenario \cite{Banerjee2018CoherentMatter}. If $\alpha$ and $m_e$ oscillate,  the resulting fractional change in $f_{atom}$ has amplitude:
\begin{equation}
\label{eq:deltafatom}
\frac{\delta f_{atom}}{f_{atom}}=2\frac{\delta\alpha}{\alpha_0}+\frac{\delta m_e}{m_{e,0}}=\Big(2g_{\gamma}+\frac{g_{e}}{m_{e,0}}\Big)\frac{\sqrt{2\rho_{\rm DM}}}{m_{\phi}}.
\end{equation}
When atoms are resonantly excited with light of stable frequency $f_L$ ($f_L\approx f_{atom}$) and in the absence  of extraneous noise sources, an observed modulation in the atomic frequency $f_{atom}$ is assumed to arise due to variations of $\alpha$ and $m_e$. In the absence of detection of such modulation, constraints can be placed on $g_{\gamma}$ and $g_{e}$. As the detected response of atoms to oscillations in $f_{atom}$ decreases at  frequencies greater than the lifetime $\tau=1/\Gamma$ of the excited state, where $\Gamma=2\pi\cdot5.2$ MHz is the natural linewidth of the 6p$^2$P$_{3/2}$ state, the measured fractional change in  $f_{atom}$  has the form: 
\begin{equation}
\label{eq:deltafatom_highfreq}
\Big(\frac{\delta f_{atom}}{f_{atom}}\Big)_{meas}=\Big(2g_{\gamma}+\frac{g_{e}}{m_{e,0}}\Big)\frac{\sqrt{2\rho_{\rm DM}}}{m_{\phi}}\Big[1+\Big(\frac{2\pi f}{\Gamma}\Big)^2\Big]^{-1/2}.
\end{equation}
\noindent The last term in Eq. (\ref{eq:deltafatom_highfreq}) is the atomic response function $h_{\rm atom}(f)$, which is $\approx1$ for $f$ which is below the cut-off $f_{\rm{cutoff2}}=\Gamma/2\pi$, and it rolls off as $1/f$ for frequencies above $f_{\rm{cutoff2}}$.

The assumption of stable frequency $f_L$ in its comparison with $f_{atom}$ requires some discussion. If $\alpha$ and $m_e$  oscillate, so does the length $L_r$ of the laser resonator, since it is a multiple of the Bohr radius $1/\alpha m_e$ \cite{Stadnik2015SearchingInterferometry,Kozlov2018CommentConstants}, leading to a modulation in $f_L$. This modulation depends on a different combination of $\delta\alpha/\alpha_0$ and $\delta m_e/m_{e,0}$, compared to that of (\ref{eq:deltafatom}), and can occur at frequencies as large as $f_{\rm{cutoff1}}=v_s/L_{r}\approx$ 50 kHz, where $L_{r}\approx$ 0.12 m is the linear dimension of the resonator and $v_s\approx$ 6000 m/s is the speed of sound in the stainless-steel-made resonator structure. At frequencies below $f_{\rm{cutoff1}}$, the induced fractional oscillation in $f_L$ has amplitude \cite{Kozlov2018CommentConstants}:
\begin{equation}
\label{eq:deltafL}
\frac{\delta f_{L}}{f_{L}}=\frac{\delta\alpha}{\alpha_0}+\frac{\delta m_e}{m_{e,0}}=\Big(g_{\gamma}+\frac{g_{e}}{m_{e,0}}\Big)\frac{\sqrt{2\rho_{\rm DM}}}{m_{\phi}}.
\end{equation}

\noindent Comparison of $f_L$ and $f_{atom}$ in the range below $f_{\rm{cutoff1}}$, therefore offers reduced sensitivity in changes of $\alpha$, while it is not sensitive to changes of $m_e$. The measured fractional variation in $f_{atom}$ in this case has amplitude: 
\begin{equation}
\label{eq:deltafatomapparent}
\Big(\frac{\delta f_{atom}}{f_{atom}}\Big)_{meas}=\frac{\delta\alpha}{\alpha_0}=g_{\gamma}\frac{\sqrt{2\rho_{\rm DM}}}{m_{\phi}},
\end{equation}

\noindent with $h_{atom}(f)$=1 here, since $f_{\rm{cutoff1}}\ll f_{\rm{cutoff2}}$. In interpreting measurements that check for rapid variations of $f_{atom}$, one has to treat the frequency regimes below and above the cut-off $f_{\rm{cutoff1}}$ differently. The Eq. (\ref{eq:deltafatomapparent}) is valid below  $f_{\rm{cutoff1}}$, while Eq. (\ref{eq:deltafatom_highfreq}) above it \footnote{
The frequency $f_L$ of the laser is referenced to the resonance frequency of an internal optical cavity, with a stabilization bandwidth of $\approx$ 5 kHz. The experiment, however, is carried out at frequencies in the range 20 kHz-100 MHz, which are higher than this additional cut-off. Therefore, the effects of FC-induced oscillations in $f_L$ arising due to the internal cavity, need not be considered. }.

\begin{figure}
\begin{subfigure}{0.5\textwidth}
  \includegraphics[width=8.5cm]{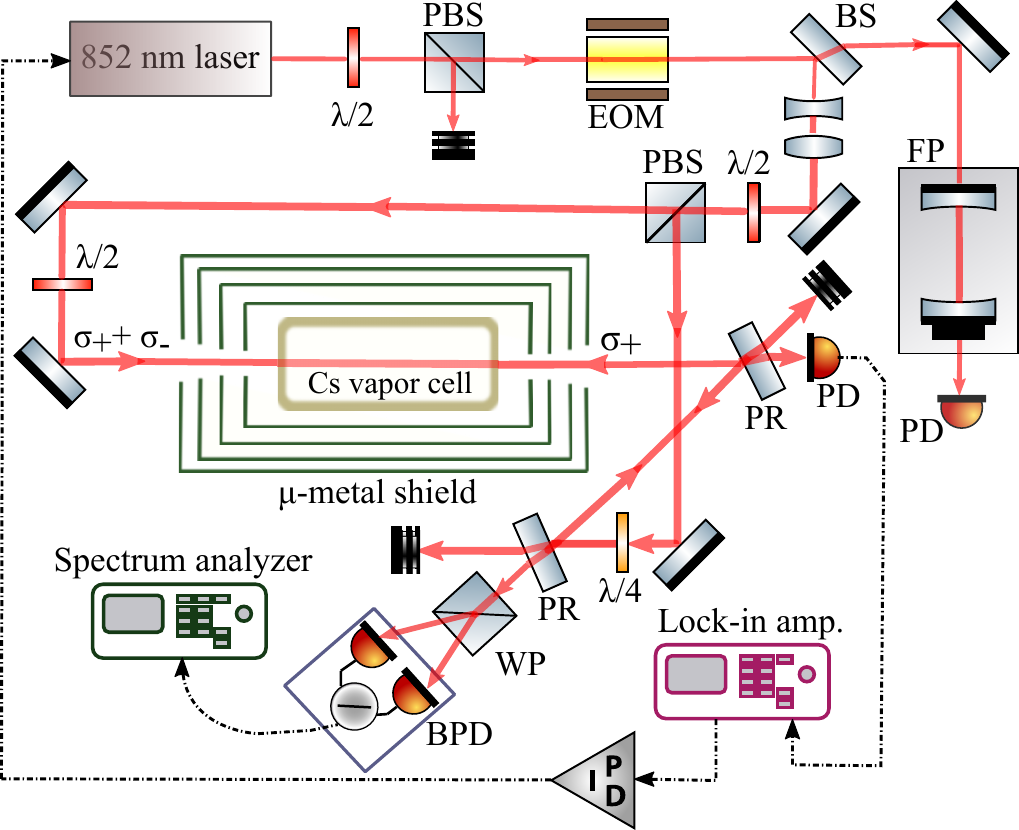}
  \caption{}
  \label{fig:apparatus}
\end{subfigure}\\
\begin{subfigure}{.5\textwidth}
  \includegraphics[width=8.5cm]{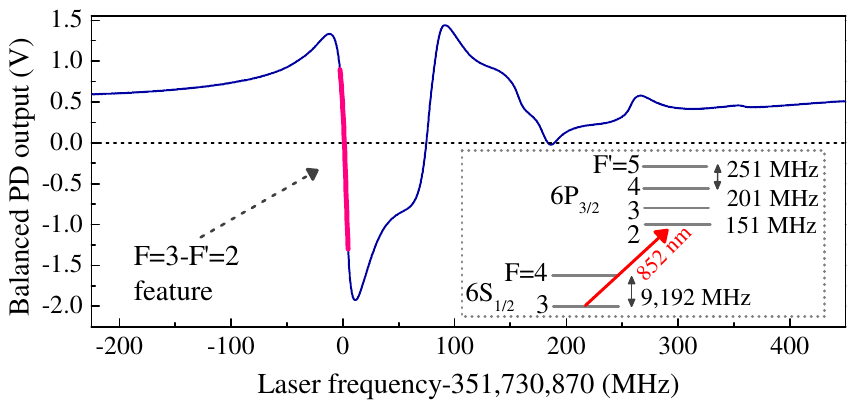}
  \caption{}
  \label{fig:polspectrum}
\end{subfigure}
\caption{\small{a) Experimental apparatus. (P)BS: (polarizing) beamsplitter; FP: Fabry-Perot interferometer; EOM: electro-optic modulator; (B)PD: (balanced) photodetector; WP: Wollaston prism; PR: partial reflector. $\lambda/2$: half-wave plate; $\lambda/4$: quarter-wave plate.  b) Polarization spectroscopy on the 6S$_{1/2}$ $F=3\rightarrow $ 6P$_{3/2}$ $F'=2,3,4$ transitions. The pink line indicates the feature employed for frequency discrimination.  Inset shows the hyperfine structure of ground 6S$_{1/2}$ state and excited 6P$_{3/2}$ state. } }
\label{fig:Apparatus-Spectrum}
\end{figure}

\textit{Experiment} - To search for fast variations in the Cs 6S$_{1/2}\rightarrow $ 6P$_{3/2}$ transition frequency, we employ polarization spectroscopy in a vapor cell \cite{Demtroder2015Laser2} (see Fig. \ref{fig:apparatus}). The 7 cm long cell is placed inside a four-layer magnetic shield and maintained at room temperature. Two counter-propagating laser beams, termed pump and probe, are overlapped inside the cell. The circularly polarized pump induces birefringence in the Cs vapor.  Analysis of the polarization of the linearly polarized probe with a balanced polarimeter yields a dispersive-shape feature against laser frequency, for each of the hyperfine components of the transition. These features have narrow widths, nearly limited by the $\approx$ 5.2 MHz natural linewidth of the transition, and serve as calibrated frequency discriminators. A typical polarization-spectroscopy signal is shown in Fig.  \ref{fig:polspectrum}. Fast changes in $f_{atom}$ will appear as amplitude oscillation in the polarimeter output. The quality factor of this oscillation is related to the coherence of the field $\phi$ of eq. (\ref{eq:phi}) and is given by $\omega/\Delta\omega\approx 2\pi/v_{\rm DM}^2\approx6\cdot10^6$, where $v_{\rm DM}=10^{-3}$ is the virial velocity of the DM field \cite{Krauss1985CalculationsDetection} (In the case of a relaxion halo, a longer coherence time is expected, resulting in a larger quality factor \cite{Banerjee2019RelaxionPhysics}.) Within the 20 kHz-100 MHz band probed in the experiment, the expected spectral width $\Delta\omega/2\pi$ of the oscillation is in the range 3 mHz - 17 Hz. 

To account for the decrease in the atomic response at frequencies above the transition linewidth, and other response non-uniformities in the apparatus, a frequency calibration is required. This is done by imposing frequency modulation on the laser light with the use of an electro-optic modulator (EOM), and comparing the amplitudes of this modulation, as measured  with the atoms and with a Fabry-Perot cavity of known characteristics that serves as a calibration reference.  

During an experiment, the laser frequency is tuned to excite atoms from the  $F=3$ hyperfine level of the ground state to the $F'=2$ level of the excited state. The output of the balanced polarimeter is measured with a spectrum analyzer (Keysight N9320B). The calibration of this analyzer for measurements of absolute power of magnitude similar to that detected in the actual experiment \mbox{($\sim$ 2 fW/ Hz)}, was checked by measurement of a signal of known power spectral density.   To produce a high resolution noise power spectrum in the 20 kHz-100 MHz range, measurements in $\approx22,000$ frequency windows are required, each of which consists of 461 bins; a bin is 10 Hz wide and corresponds to integration time of 5 ms. Approximately 22 hr is required to acquire such a spectrum. To ensure long-term frequency stability of the laser, its frequency is stabilized to the atomic resonance. This is achieved by modulating $f_L$ at 167 Hz with an amplitude of 200 kHz, and demodulation of the measured probe beam power with a lock-in amplifier provides an error signal, to which the laser frequency is stabilized with a bandwidth of 2 Hz.

\begin{figure}
\includegraphics[width=8.7cm]{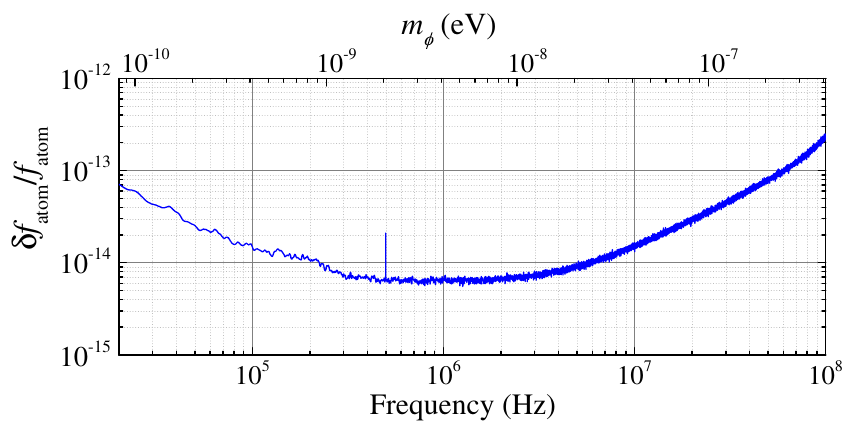}
\caption{\small{Upper bounds on the fractional modulation $\delta f_{atom}/f_{atom}$, shown at the 95\% CL. The reduced sensitivity in the range 498,330 $\pm$ 5 Hz, is due to increased apparatus noise. At frequencies below 300 kHz the sensitivity is limited by technical noise of the laser and above 5 MHz by the decaying response of the atoms. In the range 300 kHz-5 MHz, the sensitivity is nearly limited by the shot noise of the probe light that is measured with the balanced polarimeter shown in Fig.  \ref{fig:apparatus} (see supplemental material).}}
\label{fig:dvv}
\end{figure}

\textit{Data analysis} - A set of three high-resolution noise-power spectra in the 20 kHz-100 MHz range were acquired and analyzed to probe fast oscillations in $f_{atom}$. The mean and variance of each spectrum were computed in several selected frequency regions and were found to be consistent among the three spectra to within 2\%. The slope of the $F=3 \rightarrow F'=2$ feature in the polarization spectrum of Fig. \ref{fig:polspectrum}, relevant to the sensitivity in detecting oscillations in $f_{atom}$, was stable to within 6\% during the entire 66 hr long acquisition run. An averaged spectrum was computed from the three high-resolution power spectra. The sensitivity in detection of FC oscillations at a given frequency is related to the fluctuations of the noise power level in that spectrum, within the particular frequency range (see analysis in supplemental material). For each frequency bin within this range, the noise fluctuations define a global threshold (i.e. accounting for the look elsewhere effect) at the 95\% confidence level \cite{Scargle1982StudiesData}. Any peak above this threshold must be then investigated for possible detection. A number of such peaks were present in the averaged spectrum. These were checked using methods described in the supplemental material. No signal of unknown origin with power above the threshold was detected. In its absence, an upper limit is placed on possible oscillations of the frequency $f_{atom}$, which is presented in Fig. \ref{fig:dvv} at the 95\% (CL).

\begin{figure}
\includegraphics[width=8.6cm]{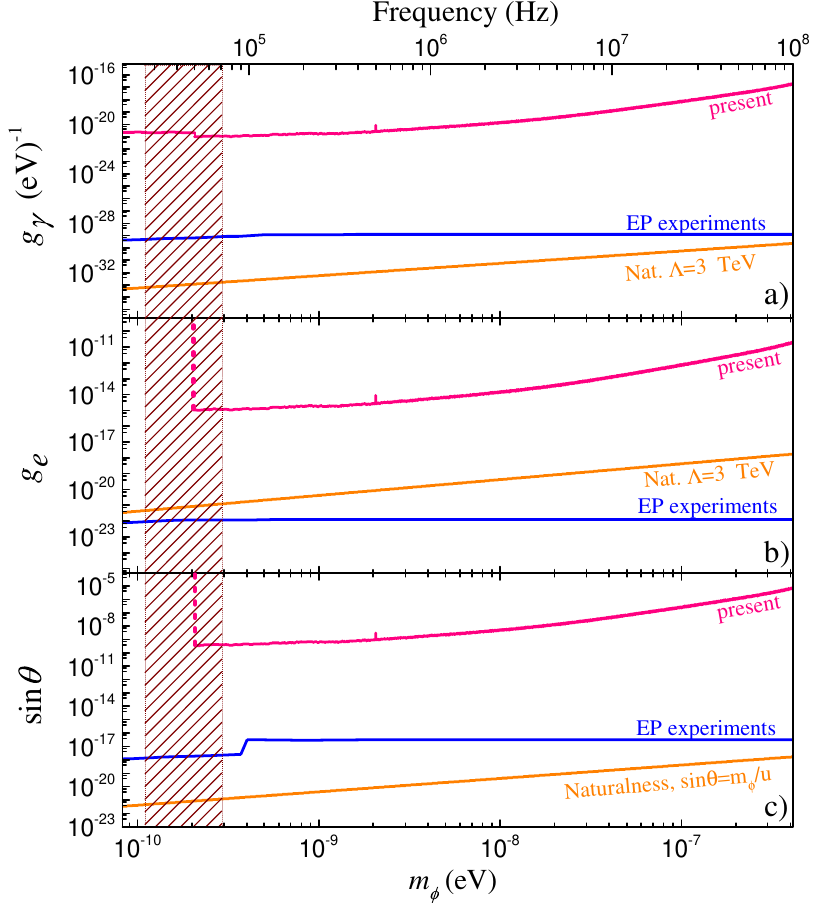}
\caption{\small{Constraints on scalar DM parameters at the 95\% CL, obtained with use of the limits on $\delta f_{atom}/f_{atom}$ shown in Fig. \ref{fig:dvv}. The bounds derived from EP experiments are from \cite{Hees2018ViolationMatter}. Constraints from the requirement for naturalness are explained in the text. The shaded area centered around the cut-off frequency of 50 kHz ($1-3\times 10^{-10}$ eV) indicates a region in which careful modeling of the laser resonator response is required to determine the transition in sensitivity to the various parameters constrained.}}
\label{fig:DMconstraints}
\end{figure}

\textit{Constraints on scalar DM couplings} - We use the obtained bounds on $\delta f_{atom} / f_{atom} $ to  constrain the parameters $g_{\gamma}$ and $g_{e}$ of Eqs (\ref{eq:alphaVar}) and (\ref{eq:mVar}). With the assumption that DM-induced  oscillations in $f_a$ arise solely due to either the coupling to the photon or to the electron, we set bounds on the corresponding coupling constants, and present these in Fig. \ref{fig:DMconstraints}a and Fig. \ref{fig:DMconstraints}b. In the same plots, corresponding limits derived from analysis \cite{Hees2018ViolationMatter} of results of EP experiments, as well as limits derived from naturalness are also shown. In the case of a scalar field $\phi$, naturalness requires that radiative corrections to the mass $m_{\phi}$, arising due to its interactions, be smaller than the mass itself \cite{Arvanitaki2016SoundDetectors, Graham2016DarkAccelerometers}. In the present work, where a DM field that has scalar couplings to SM matter is considered, this requirement leads to the constraints: $\vert g_e\vert< 4\pi m_{\phi}/\Lambda$, $\vert g_{\gamma}\vert<16\pi m_{\phi}/\Lambda^2$, where $\Lambda$ is the cut-off scale for the Higgs mass \cite{Banerjee2018CoherentMatter}.

To obtain the bounds of Fig. \ref{fig:DMconstraints}a and \ref{fig:DMconstraints}b, $g_{\gamma}$ and $g_{e}$ were treated independently. Within the relaxion DM model \cite{Banerjee2018CoherentMatter}, however, this assumption is not valid. These couplings are related; both acquire values dependent on the relaxion-Higgs mixing, which is parametrized in terms of a mixing angle $\theta $. For the range of  mass $m_{\phi}$ probed in this work ($10^{-10}$-$10^{-6}$ eV), the contribution of the relaxion coupling to the electron is expected to dominate the oscillations in $f_a$ \cite{Banerjee2018CoherentMatter}. One can therefore assume that $\delta f_{atom}/f_{atom}\approx\delta m_e/m_e$ in Eq. (\ref{eq:deltafatom}) and employ the defining relation between $g_e$ and the mixing angle $\theta$, to constrain $\theta$ within the investigated $m_{\phi}$ region. The parameter $g_e$ is given by \cite{Banerjee2018CoherentMatter} :
\begin{equation}
\label{eq:ge-theta}
g_e=Y_e\sin\theta,
\end{equation}

\begin{figure}
\includegraphics[width=8.7cm]{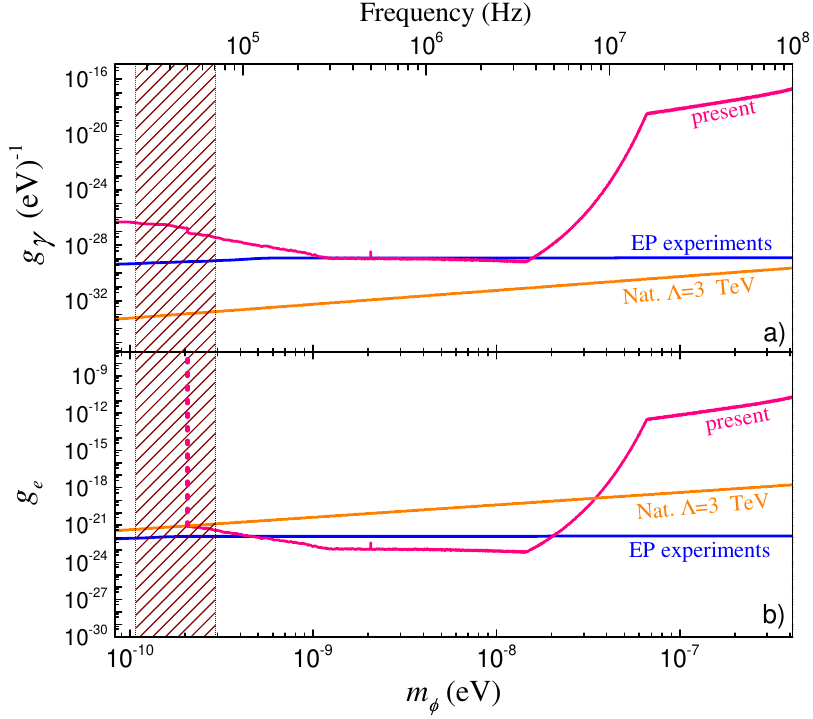}
\caption{\small{Bounds on the parameters $g_{\gamma}$, $g_e$ with consideration to a relaxion halo gravitationally bound by earth. The DM density assumed within this scenario is computed in \cite{Banerjee2019RelaxionPhysics}. The shaded area is the same as that in Fig. \ref{fig:DMconstraints}. }}
\label{fig:Haloconstraints}
\end{figure}

\noindent where $Y_e$ is the Higgs-electron Yukawa coupling, for which the accepted value within the SM is: $Y_e\approx 2.9\cdot 10^{-6}$ \cite{Altmannshofer2015ExperimentalElectrons}. We show the obtained bounds on $\sin\theta$ in Fig. \ref{fig:DMconstraints}c, along with corresponding bounds placed from EP experiments, and by the requirement to maintain naturalness. Within the relaxion DM framework, this requirement results in the constraint $\sin\theta\leq m_{\phi}/v$ \cite{Frugiuele2018RelaxionColliders} where $v=246$ GeV is the Higgs vacuum expectation value. 

An enhancement in the amplitude of FC oscillations is expected in the presence of an astronomical-scale DM object around the earth or in its vicinity. Such an enhancement would occur due to an increase in the local DM density $\rho_{\rm DM}$ [see eqns (\ref{eq:phi}), (\ref{eq:alphaVar}), (\ref{eq:mVar})]. Searches for transient variations of $\alpha$ using a network of GPS satellites \cite{Roberts2017SearchSatellites} and a terrestrial network of remotely located atomic clocks \cite{Wcisoetal.2018NewClocks}, have provided constraints on topological DM. Here we consider the scenario of an earth-bound relaxion halo, examined in \cite{Banerjee2019RelaxionPhysics}. We make use of the computed DM density $\rho_{\rm DM}$ \cite{Banerjee2019RelaxionPhysics} at the surface of the Earth, to provide more stringent constraints on the couplings $g_{\gamma}$  and $ g_{e}$ than these shown in Fig. \ref{fig:DMconstraints} (conditional on the existence of the relaxion halo). These tighter bounds are presented in Fig. \ref{fig:Haloconstraints}. In the presence of the halo, the enhancement in $\rho_{DM}$ is expected to be mostly pronounced in the mass range $10^{-12}$-$10^{-8}$ eV. We note that the corresponding frequency regime $10^4$-$10^8$ Hz has been out of reach for most of the laboratory searches for variations of FC, which with the exception of the recent work \cite{Aharony2019ConstrainingDecoupling}, have been mostly sensitive to frequencies below 1 Hz.

\textit{Discussion and outlook} - The obtained constraints on light DM scalar interactions with SM matter extend the frequency range investigated with atomic probes to the $10^8$ Hz, a regime not previously searched directly. More stringent bounds on the scalar coupling to the photon and the electron exist within the 20 kHz-100 MHz range of this work. Such constraints, however, are derived from EP and FF experiments, which are not directly sensitive to rapid oscillations of FC, as is the method demonstrated here.

The sensitivity in detection of DM-induced FC variations for given bound on $\delta f_{atom}/f_{atom}$ is inversely proportional to the frequency/mass probed [see Eq. (\ref{eq:deltafatom})]. Use of atomic clocks is therefore advantageous, in that these probes typically search the sub-Hz regime, with resulting DM constraints competing against those provided by EP and FF studies. Without assuming any enhancement in the local DM density due to the Earth, our polarization spectroscopy scheme is unlikely to approach a sensitivity level comparable to that offered by EP/FF searches because of the higher frequencies being probed. However, searching for rapid FC oscillations in the rf-band allows to access a frequency range, which as discussed, might provide enhanced sensitivity in detection of FC variations within scenarios of an earth-bound DM-halo.  As seen in Fig. \ref{fig:Haloconstraints}, the bounds from polarization spectroscopy are at regions better than those set by EP results.  Improvements in detection of modulation in the Cs energy levels  involved in the experiment, would enable a search deeper into the parameter space of $g_{\gamma}$ and $g_{e}$.  

Several apparatus upgrades are under way. The primary improvement involves a change in the scheme employed to obtain the frequency spectrum of the polarization-spectroscopy signal. The spectrum analyzer currently used will be replaced with fast data-acquisition electronics with the ability to perform real-time high-resolution spectral analysis, thereby increasing the effective integration time drastically.  An optimization of the signal-to-noise ratio in the polarization spectroscopy setup is also being explored. These improvements, combined with a longer integration time of up to $\sim 10^4$ hr, should result in a sensitivity enhancement in excess of $10^3$. This level of improvement is sufficient to explore scalar interactions between DM and SM matter with a sensitivity better than that of other methods in a significant fraction of the parameter space accessible by polarization spectroscopy.\\

\section*{Acknowledgements}
We are grateful to M. G. Kozlov, V. V. Flambaum, V. Dzuba and Y. Stadnik for fruitful discussions. We acknowledge technical support from  A. Brogna, M. Schott, T.H. Lin and A. D{\"u}dder. The work is supported by the European Research Council (ERC) under the European Unions Horizon 2020 research and innovation programme (Dark-OST, grant agreement No 695405), and the DFG Reinhart Koselleck project. The work of GP is supported by grants from the BSF, ERC, ISF; the work of RO and GP is jointly supported by the Minerva Foundation, and the Segre Research Award. The work of RO is supported by the ISF, ERC and the Israeli Ministry of Science and Technology.

\bibliographystyle{apsrev4-1}
\bibliography{Mendeley}

\section*{Supplemental material}

\subsection{Data analysis scheme}
\begin{figure}
\includegraphics[width=8.6cm]{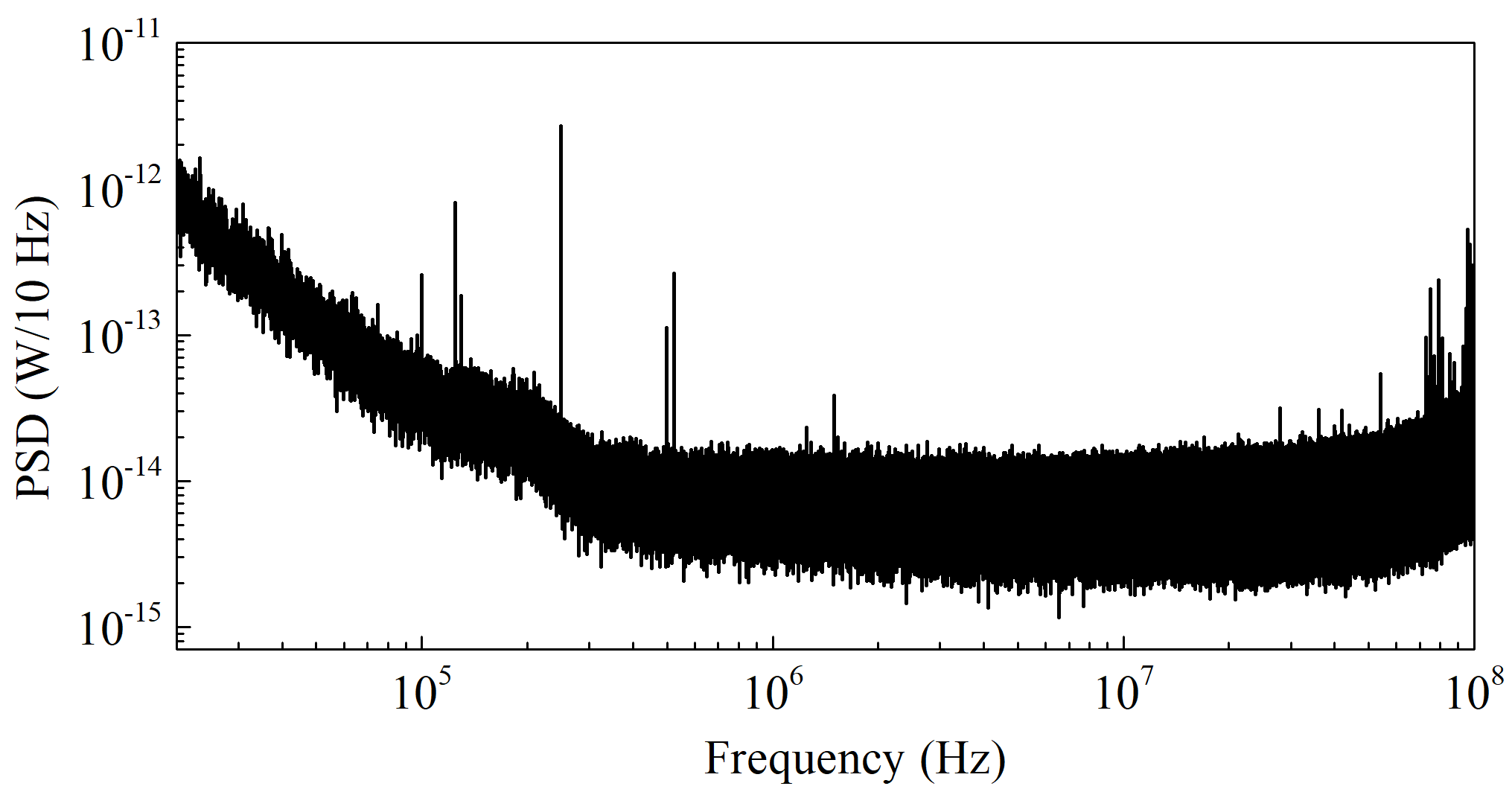}
\caption{Power spectral density of the balanced photodetector output, calculated as the average of three 22 hr-long data acquisition runs performed in this experiment. The averaged spectrum consists of $\sim10^7$ data points, each of which corresponds to the measured power within a $10$ Hz bin.}
\label{fig:PowerSpectrum}
\end{figure}

  The power spectral density (PSD) analyzed to impose the DM constraints reported in this work  is computed by averaging three PSDs in the range 20 kHz-100 MHz (see \textit{``Data analysis"} section), and is shown in Fig. \ref{fig:PowerSpectrum}.
  
   
  A first step in the data analysis consists in identifying obvious outliers and checking for possible dark-matter (DM) induced signal detection.
  A number of such peaks were present in the averaged spectrum. 
  Those at frequencies lower than $1.1$ MHz were all attributed to technical laser noise (either frequency or amplitude noise), or detector background noise.
  Their origin  was identified via one or more of a few different methods.  
  DM-candidate signals were checked in the laser-intensity spectrum with a photodetector. 
  The power of candidate peaks was compared with the laser frequency tuned on or off the $F = 3 \rightarrow F' = 2$ resonance.
  In addition, we used a FP resonator to discriminate frequency from amplitude noise contributions to the  candidate peaks.
  Finally, data were acquired with a second Ti:Sapphire laser and an external-cavity diode laser.
  These lasers have different technical-noise spectra compared to those of the primary laser system.
  
  All but one candidate peaks at frequencies below $1.1$ MHz were completely accounted for, i.e. their power was measured and subtracted from the spectrum, and it was confirmed that there is no statistically significant residual power at their respective positions.
  
  All spurious signals at frequencies above $1.1$ MHz have an origin which is different from  the origins of the peaks below  $1.1$ MHz. They were all found to arise from rf-pickup in the apparatus, such as, for example, due to radio broadcasting.
  Subtracting the spectra with the laser tuned on and off the atomic resonance allowed us to exclude a possible DM-induced signal at the respective peak positions.
  
  Thus, with the exception of a single peak discussed below, all DM-candidate features exceeding a $\approx5.7\sigma$ level could be subtracted and were ignored in the subsequent analysis. That is, 
  the spectral regions where the peaks were subtracted were treated the same way as the rest of the spectrum in the subsequent data analysis. 
  In this analysis, the amplitude spectral density (ASD - in units of V/$\sqrt{10\:\rm{Hz}}$) shown in Fig.  \ref{fig:AmplitudeSpectrum}, was used. 

  A single peak, measured at $498,330\pm5$ Hz could not be completely subtracted, due to the imperfect ability to accurately measure its frequency and amplitude contributions to the noise, that are present in the frequency and intensity spectra of the laser light, respectively. As this spurious peak is present in noise measurements which are insensitive to DM, we excluded the possibility that it arises due to DM. 
  In contrast to the other peaks exceeding the threshold which we were able to identify as associated with laser technical noise and subtract without loss of sensitivity, we could not do a reliable subtraction in this case, which has resulted in a reduced sensitivity at this frequency.
  Indeed, after subtracting the estimated $498,330\pm5$ Hz peak power, the residual power at this frequency still differs by a factor of two from the local mean power.
  Therefore, this reduces the experimental sensitivity at this particular frequency by a factor of two.

\begin{figure}
\includegraphics[width=8.6cm]{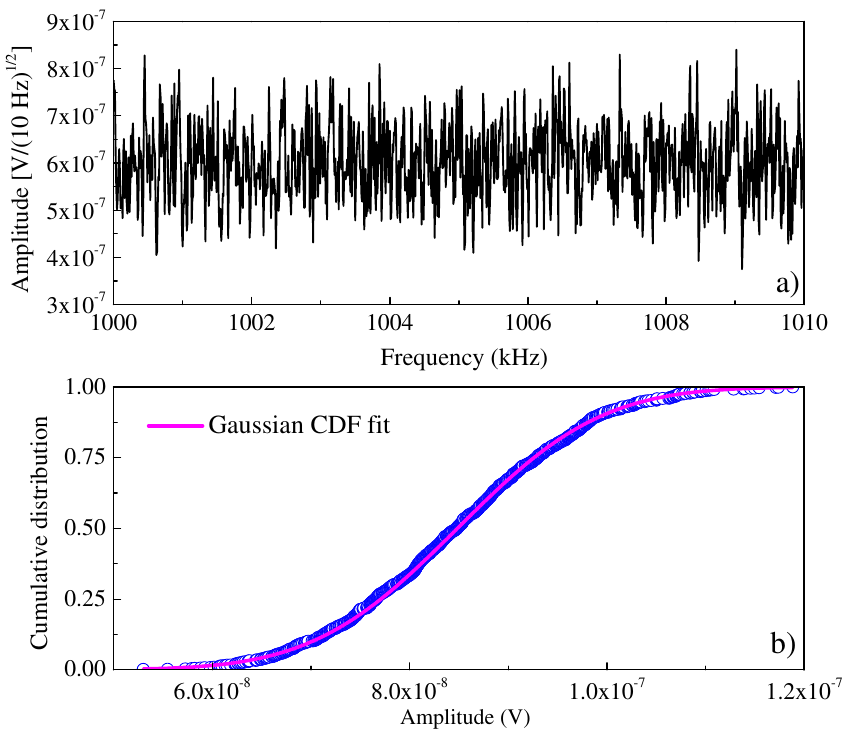}
\caption{a): Amplitude noise spectrum calculated from the data of Fig. \ref{fig:PowerSpectrum} within a 10 kHz-wide frequency window. b): Corresponding CDF of noise, and Gaussian CDF fit to these data.} 
\label{fig:CDF}
\end{figure}

\begin{figure}
\includegraphics[width=8.6cm]{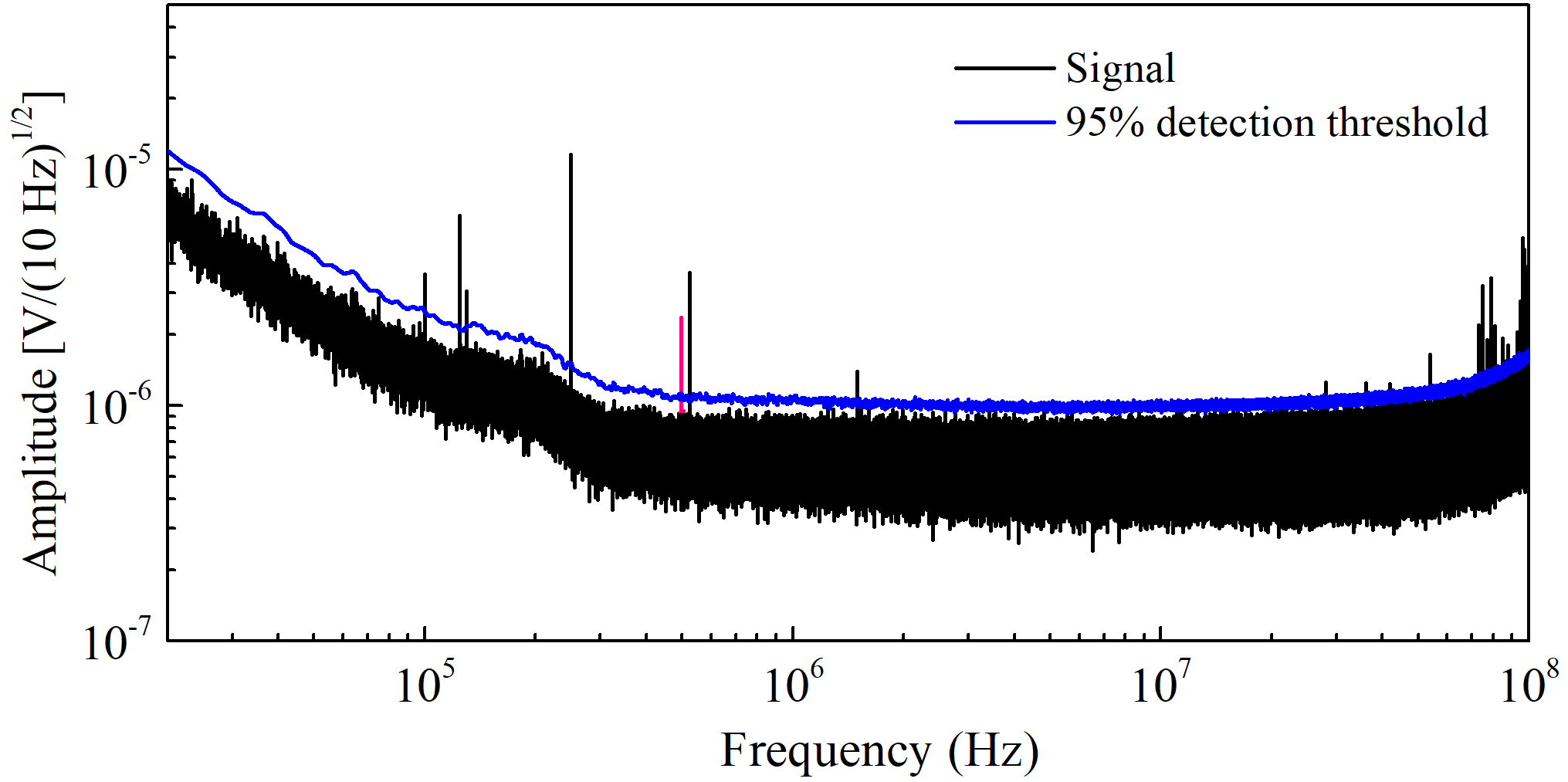}
\caption{Amplitude spectrum and detection threshold shown at the $95\%$ level.  The peak in magenta color indicates a spurious signal at $498,330\pm5$ Hz, in which frequency the sensitivity in constraining a DM-induced signal is decreased (see text). }
\label{fig:AmplitudeSpectrum}
\end{figure}

  The ASD noise is used to define a voltage detection threshold, $V_{\rm{th}}$, such that if, at a given frequency, a signal is measured with amplitude $V > V_{\rm{Th}}$,  then it had a probability \mbox{$p_0=5\%$} to be induced by noise fluctuations (corresponding to the $95\%$ confidence level).
  One can write this statement as:
   \begin{align}
        p_0  &=    \mathcal{P}(V > V_{\rm{th}})  \\
             & =  1-\mathcal{P}(V < V_{\rm{th}})~, \label{eq2}
   \end{align}
  where $\mathcal{P}(V < V_{\rm{Th}})$, also known as the cumulative distribution function (CDF), denotes the probability of measuring $V < V_{\rm{th}}$ if $V$ is noise-induced.
  In this experiment the ASD noise was identified to be Gaussian distributed (see for example the noise ASD CDF in a $10$~kHz window around $1$ MHz in Fig.~\ref{fig:CDF}b).

  Inserting the analytical expression for the Gaussian CDF yields:
  \begin{align}
      V_{\rm{th}} &= \sqrt{2}~\rm{erf} ^{-1}\Big\{1-2p_0\Big\} \sigma + \mu \\ 
      &\approx 1.64 \sigma + \mu~, \label{CDFfit}
  \end{align}
  where $\mu$ and $\sigma$ are the mean value and standard deviation of the Gaussian distribution and $\rm{erf}$ is the error function.
  In order to estimate $\mu$ and $\sigma$, we fit the ASD CDF in  $10$~kHz windows to a Gaussian CDF. 
  The values for $\mu$ and $\sigma$ from the fit are used to evaluate Eq. \eqref{CDFfit} and determine $V_{\rm{th}}$. An example of such a fit is shown in Fig. \ref{fig:CDF}b. 
  Repeating this operation in the 20 kHz-100 MHz investigated frequency range yields the parameters $\mu$ and $\sigma$, and therefore  $V_{\rm{th}}$, in the entire bandwidth. 
 
  Before proceeding to search for possible DM-induced peaks exceeding $V_{\rm{th}}$, we apply a statistical penalty to account for the fact that many frequency data points are inspected, thereby leading to a large number of noise-fluctuation-induced peaks above the threshold ($\sim 5\%$ at the $95\%$ confidence level).
  This is commonly referred to as the look-elsewhere effect.
  In practice, we raise Eq. \eqref{CDFfit} to the power of the number of inspected frequencies~\cite{Scargle1982StudiesData}. 
  The  Eq.~\eqref{eq2} now reads
  
  \begin{align}
        p_0   & =   1-\mathcal{P}(V < V_{\rm{th}})^N~.\label{p-value}
   \end{align}
   
  For $N\approx10^7$ frequencies, this yields a corrected detection threshold of:
  
   \begin{align}
      V_{\rm{th}} &= \sqrt{2}~\rm{erf} ^{-1}\Big\{2(1-p_0)^{1/N}-1\Big\}\sigma + \mu \\
      &\approx 5.73 \sigma + \mu~.
  \end{align}
  
 \noindent  The calculated detection threshold is shown superimposed with the ASD in Fig. \ref{fig:AmplitudeSpectrum}.
  No peaks above the threshold have been detected in the censored ASD.
  In such a case, we place upper bounds on the possible DM-induced modulation $\delta f_{atom}/f_{atom}$. These bounds are related to the fluctuations of the noise around its mean value, and to the slope \textit{S} of the spectral feature shown  in Fig. 1b of the main manuscript: 
  \begin{equation}
      \frac{\delta f_{atom}}{f_{atom}}= \frac{1}{f_{atom}}\frac{V_{\rm{th}}-\mu}{S}=\frac{1}{f_{atom}}\frac{5.73\sigma}{S}~.
      \label{eqfracmod}
  \end{equation}
Application of Eq. (\ref{eqfracmod}) yields the limits presented in Fig. 2 of the main paper, by incorporating an overall calibration of the apparatus frequency response, which is described in the section \textit{``Experiment}". 

\subsection{Apparatus sensitivity limits}

The constraints on a possible DM-induced oscillation of the Cs energy levels, are determined by the apparatus sensitivity. As seen in Fig. 2 of the main paper, the sensitivity at low frequencies (up to 300 kHz) has a 1/\textit{f} profile. Frequency noise of the Ti:Sapphire laser is the primary contributor to this profile. At frequencies higher than the $\approx$ 5.2 MHz natural linewidth of the atomic transition, the sensitivity is primarily limited by the decaying response of atoms [see Eq. (5) in the main paper]. The intermediate frequency regime (300-kHz-5 MHz) is arguably the most interesting with regard to DM searches within the relaxion-halo model, as the resulting DM overdensity in this case is maximal within this intermediate frequency region. The relative frequency sensitivity of $\approx6\times10^{-15}$ achieved within this region is nearly limited by the shot noise of the light measured with the Thorlabs PBD415A balanced photodetector (see section ``\textit{Experiment}" of main paper). This shot noise can be estimated from the approximate power of 220 $\rm{\mu}$W at 852 nm that is measured with the detector, which has a responsivity of $0.53$ A/W at the particular wavelength and an effective transimpedance gain of 2.5$\times10^4$. The estimated shot noise corresponds to $\approx3\times10^{-7}$ V/$\sqrt{10\:\rm{Hz}}$. In comparison,  the overall noise  around the 1 MHz range is  $\approx6\times10^{-7}$ V/$\sqrt{10\:\rm{Hz}}$, as seen in Fig. \ref{fig:CDF}a.  (The detector background noise contributes by a measured $\approx2.8\times10^{-7}$ V/$\sqrt{10\:\rm{Hz}}$ in the same frequency range.) Thus, the overall noise in our experiment is within  a factor of two of the shot noise.

\end{document}